\documentclass[twocolumn,showpacs,preprintnumbers,amsmath,amssymb]{revtex4}
\usepackage{graphicx}
\usepackage{epsfig}
\usepackage{dcolumn}
\usepackage{bm}
\def\nA{nucleon-nucleus\ }
\def\pA{proton-nucleus\ }

\begin{document}
\preprint{Draft for Phys. Rev. C}
\title{Probing the isovector transition strength of the low-lying nuclear excitations
 induced by inverse kinematics proton scattering}
\author{Dao T. Khoa}\email{khoa@vaec.gov.vn}
\affiliation{Institute for Nuclear Science {\rm \&} Technique, VAEC,
 P.O. Box 5T-160, Nghia Do, Hanoi, Vietnam.}
\date{\today}
\begin{abstract}
A compact approach based on the folding model is suggested for the determination
of the isoscalar and isovector transition strengths of the low-lying ($\Delta
S=\Delta T=0$) excitations induced by inelastic proton scattering measured with
exotic beams. Our analysis of the recently measured inelastic $^{18,20}$O+p
scattering data at $E_{\rm lab}=30$ and 43 MeV/nucleon has given for the first
time an accurate estimate of the isoscalar $\beta_0$ and isovector $\beta_1$
deformation parameters (which cannot be determined from the (p,p') data alone by
standard methods) for 2$^+_1$ and $3^-_1$ excited states in $^{18,20}$O. Quite
strong isovector mixing was found in the 2$^+_1$ inelastic $^{20}$O+p scattering
channel, where the strength of the isovector form factor $F_1$ (prototype of the
Lane potential) corresponds to a $\beta_1$ value almost 3 times larger than
$\beta_0$ and a ratio of nuclear transition matrix elements $M_n/M_p\simeq 4.2$.
\end{abstract}
 \maketitle

Although the isospin dependence of the \nA optical potential, known by now as
Lane potential \cite{La62}, has been studied since a long time, very few
attempts have been made to study the isospin dependence of the transition
potential for \emph{inelastic} scattering. The neutron and proton contributions
to the structure of the low-lying nuclear excitations are known to be quite
different \cite{Be83}, and the inelastic nuclear form factor contains,
therefore, an isospin dependence which determines the degree of the \emph{
isovector} mixing in the inelastic scattering channel that induces the
excitation.

In general, the isospin-dependent potential is proportional to the product of
the projectile and target isospins ($\bm{T}_p\bm{T}_t$). For the heavy ions,
this term has been shown \cite{Kh96} to be negligible and the scattering cross
section is mainly determined by the isoscalar term. Situation is different in
the \nA case where the optical potential can be written in terms of the
isoscalar (IS) and isovector (IV) components \cite{La62} as
\begin{equation}
 U(R)=U_0(R)\pm\varepsilon U_1(R), \ \varepsilon=(N-Z)/A, \label{e1}
\end{equation}
where the + sign pertains to incident neutron and - sign to incident proton. The
strength of the Lane potential $U_1$ is known from (p,p) and (n,n) elastic
scattering and (p,n) reactions studies, to be around 30-40\% of the $U_0$
strength. In many cases, inelastic \nA scattering cross section can be
reasonably well described, in the distorted-wave Born approximation (DWBA) or
coupled channel formalism by a collective-model prescription, where the
inelastic form factor $F$ is obtained by `deforming' the optical potential
(\ref{e1}) with a scaling factor $\delta$ known as the nuclear deformation
length
\begin{equation}
 F(R)=\delta\frac{dU(R)}{dR}=\delta_0\frac{dU_0(R)}{dR}\pm\varepsilon\
 \delta_1\frac{dU_1(R)}{dR}. \label{e2}
\end{equation}
The explicit knowledge of $\delta_0$ and $\delta_1$ would give us vital
structure information about the IS and IV transition strengths of the excitation
under study. There are only two types of experiment that might allow one to
determine $\delta_0$ and $\delta_1$ using prescription (\ref{e2}):

i) (p,n) reaction leading to the \emph{excited} isobar analog state. It was
shown, however, that the two-step mechanism usually dominates this process and
the calculated cross sections were insensitive to $\delta_1$ values \cite{Fi79}.

ii) Another way is to extract $\delta_{0(1)}$ from the (p,p') and (n,n') data
measured at about the same incident energy and exciting the same state of the
target \cite{Fi79,Gr80}. Since $\varepsilon U_1/U_0$ is only about few percent,
the uncertainty of such method can be quite large. Moreover, the most
interesting data are now being measured with the secondary (unstable) beams and
(given the beam intensities much weaker than those of stable beams) it is
technically not feasible to perform simultaneously (p,p') and (n,n')
measurements (in the inverse kinematics) with those beams.

From a theoretical point of view, the form factor (\ref{e2}) has been shown to
have inaccurate radial shape which tends to underestimate the transition
strength, especially, for high-multipole excitations induced by inelastic
heavy-ion scattering \cite{Be95,Kh00}. Since the nuclear deformation is directly
linked to the `deformed' shape of the excited nucleus, instead of `deforming'
the optical potential (\ref{e2}), we build up the proton and neutron transition
densities of a $2^{\lambda}$-pole excitation ($\lambda\ge 2$) using
Bohr-Mottelson (BM) prescription \cite{Bo75} separately for protons and neutrons
\begin{equation}
 \rho^\tau_\lambda(r)=-\delta_\tau\frac{d\rho^\tau_{\rm g.s.}(r)}{dr},\ {\rm with}
 \ \tau=p,n. \label{e3}
\end{equation}
Here $\rho^\tau_{\rm g.s.}(r)$ are the proton and neutron ground state (g.s.)
densities and $\delta_\tau$ are the corresponding deformation lengths. Given the
explicit proton and neutron transition densities and an effective
isospin-dependent nucleon-nucleon (NN) interaction, one obtains from the folding
model \cite{Kh02} the inelastic \pA form factor (in terms of IS and IV parts) as
\begin{equation}
 F(R)=F_0(R)-\varepsilon F_1(R), \label{e4}
\end{equation}
where $F_0(R)=V_{IS}(R)$ and $F_1(R)=-V_{IV}(R)/\varepsilon$. The explicit
formulas of $V_{IS(IV)}$ are given in Ref.~\cite{Kh02}. One can see that
$F_1(R)$ is prototype of the Lane potential for inelastic scattering. In both
elastic and inelastic channels, $V_{IS}$ and $V_{IV}$ are determined by the sum
($\rho_n+\rho_p$) and difference ($\rho_n-\rho_p$) of the neutron and proton
densities \cite{Kh02}, respectively. It is, therefore, natural to represent the
IS and IV parts of the nuclear density as
\begin{equation}
 \rho^{0(1)}_{\lambda(g.s.)}(r)=\rho^n_{\lambda(g.s.)}(r)
 \pm\rho^p_{\lambda(g.s.)}(r).
 \label{e5}
\end{equation}
On the other hand, one can generate using the same BM prescription the IS and IV
transition densities by deforming the IS and IV parts of the nuclear g.s.
density
\begin{equation}
 \rho^{0(1)}_\lambda(r)=-\delta_{0(1)}\frac{d[\rho^n_{\rm g.s.}(r)\pm
 \rho^p_{\rm g.s.}(r)]}{dr}.
 \label{e6}
\end{equation}
The explicit expressions for the IS and IV deformation lengths are easily
obtained from Eqs.~(\ref{e5}) and (\ref{e6}), after some integration in parts,
as
\begin{equation}
 \delta_{0}=\frac{N<r^{\lambda-1}>_n\delta_n+Z<r^{\lambda-1}>_p\delta_p}
{A<r^{\lambda-1}>_A},
 \label{e6s}
\end{equation}
\begin{equation}
 \delta_{1}=\frac{N<r^{\lambda-1}>_n\delta_n-Z<r^{\lambda-1}>_p\delta_p}
{N<r^{\lambda-1}>_n-Z<r^{\lambda-1}>_p}.
 \label{e6v}
\end{equation}
The radial momenta $<r^{\lambda-1}>_{n,p,A}$ are taken over the neutron, proton
and total g.s. densities, respectively,
\begin{equation}
 <r^{\lambda-1}>_x=\int_0^\infty\rho^x_{\rm g.s.}(r)r^{\lambda+1}dr\Big/
 \int_0^\infty\rho^x_{\rm g.s.}(r)r^2dr.
\end{equation}
The transition matrix element associated with a given component of nuclear
transition density is
\begin{equation}
 M_x=\int_0^\infty \rho^x_{\lambda}(r)r^{\lambda+2}dr.
\label{e7}
\end{equation}
Realistic estimate for $M_n/M_p$ or $M_1/M_0$ should give important information
on the IS and IV transition strengths
\begin{equation}
\frac{M_n}{M_p}=\frac{N<r^{\lambda-1}>_n\delta_n}{Z<r^{\lambda-1}>_p\delta_p},
 \label{e7a}
\end{equation}
\begin{equation}
 \frac{M_1}{M_0}=\frac{(N<r^{\lambda-1}>_n-Z<r^{\lambda-1}>_p)\delta_1}
 {(A<r^{\lambda-1}>_A)\delta_0}.
 \label{e7b}
\end{equation}
It is useful to note that the ratios of transition matrix elements in the two
representations are related by
\begin{equation}
 M_n/M_p=(1+M_1/M_0)/(1-M_1/M_0).
 \label{e7c}
\end{equation}
If one assumes that the excitation is purely \emph{isoscalar} and the neutron
and proton densities have the same radial shape, scaled by the ratio $N/Z$, then
$\delta_n=\delta_p=\delta_0=\delta_1$,
\begin{equation}
 \frac{M_n}{M_p}=\frac{N}{Z}\ \ {\rm and}\
 \ \frac{M_1}{M_0}=\frac{N-Z}{A}=\varepsilon. \label{e8}
\end{equation}
Therefore, any significant difference between $M_n/M_p$ and $N/Z$ (or between
$M_1/M_0$ and $\varepsilon$) would directly indicate an isovector mixing effect.

Note that if one neglects the difference between different radial momenta
$<r^{\lambda-1}>_x$ then expressions (\ref{e7a}) and (\ref{e7b}) are reduced to
those used earlier for the `experimental' determination of $M_n/M_p$ \cite{Be83}
and $M_1/M_0$ \cite{Gr80} ratios
\begin{equation}
 \frac{M_n}{M_p}=\frac{N~\delta_n}{Z~\delta_p}\ \ {\rm and}\
 \ \frac{M_1}{M_0}=\frac{(N-Z)~\delta_1}{A~\delta_0}. \label{e9}
\end{equation}

We further choose the proton deformation length $\delta_p$ so that the
\emph{measured} electric transition rate is given by
$B(E\lambda\uparrow)=e^2|M_p|^2$. As a result, the only free parameter to be
determined from the DWBA fit to the inelastic scattering data is the neutron
deformation length $\delta_n$ if the experimental $B(E\lambda\uparrow)$ value is
known (from, e.g., $\gamma$-decay strength). Other transition matrix elements
and deformation parameters can be directly obtained from $\delta_{p(n)}$ using
Eqs.~(\ref{e3})-(\ref{e7b}). This is the main advantage of our approach compared
to the standard analysis using simple prescription (\ref{e2}).

\begin{figure*}[htb]\vspace*{-2cm}
\begin{minipage}[t]{8.5cm}
\hspace*{-1cm}
\mbox{\epsfig{file=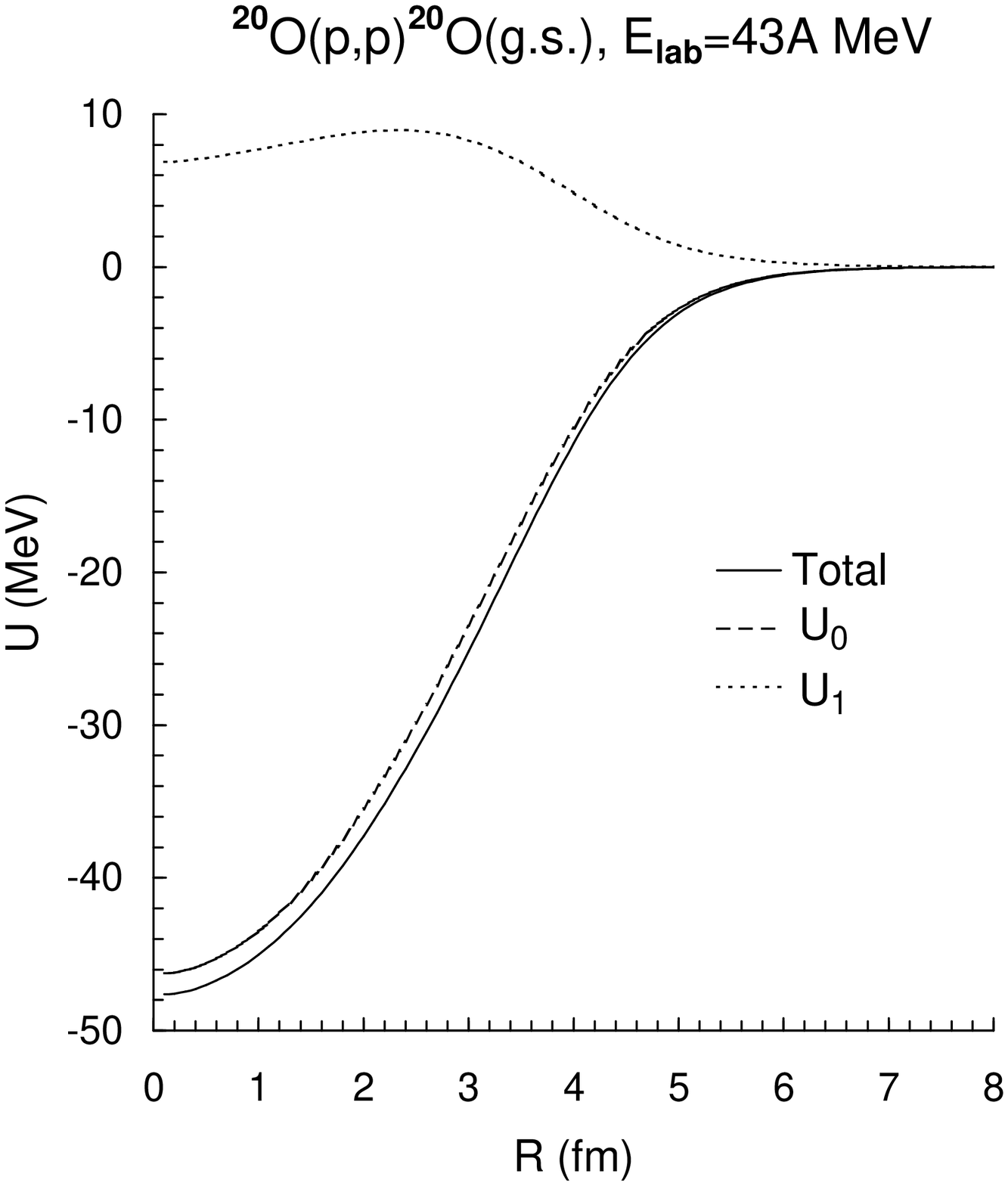,height=10cm}}
\end{minipage}
\hspace{\fill}
\begin{minipage}[t]{8.5cm}
\hspace*{-2cm}
\mbox{\epsfig{file=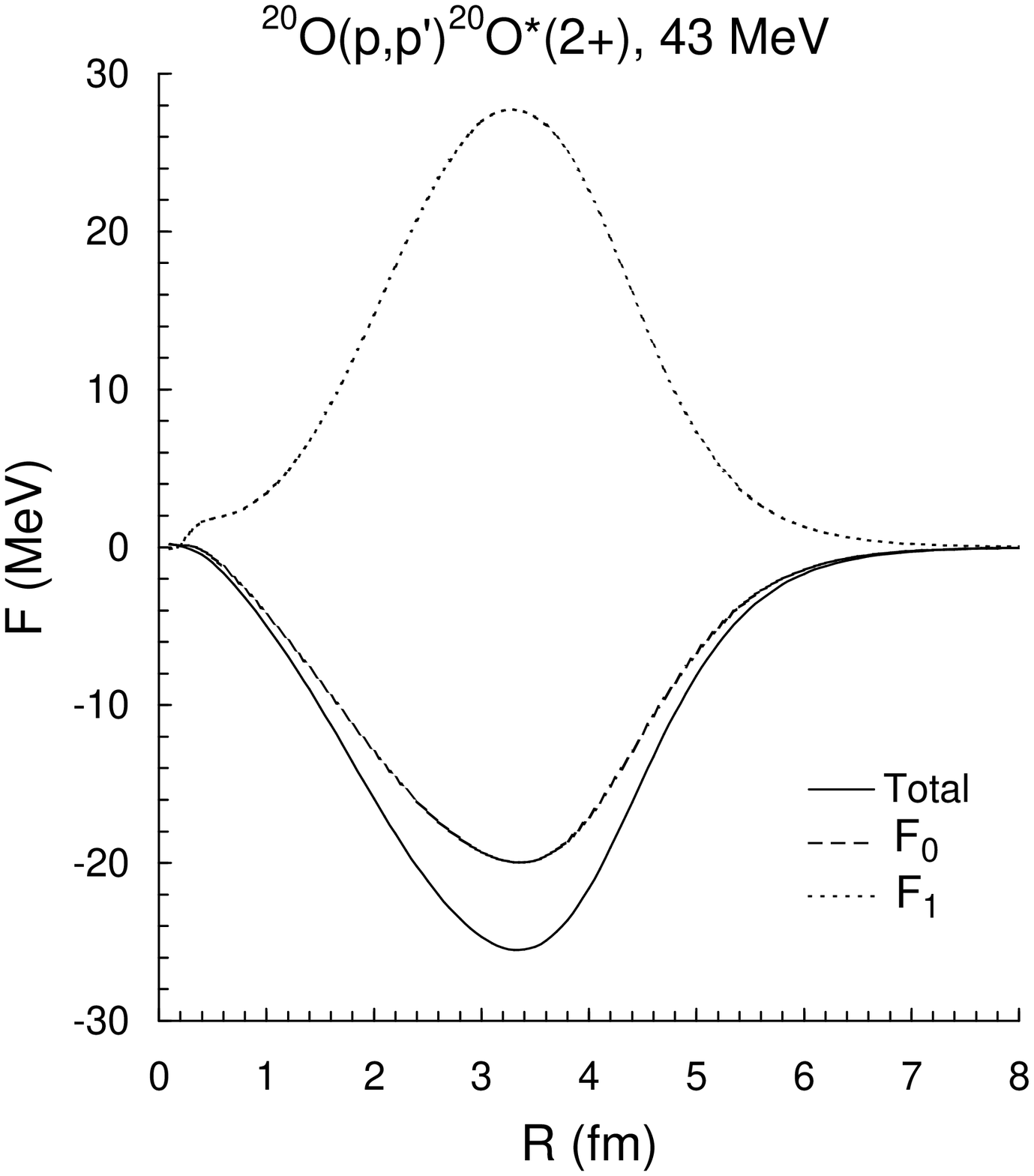,height=10cm}}
\end{minipage}
\vspace*{-2cm} \caption{Real folded optical potential (left panel) and
2$^+$-inelastic form factor (right panel) for $^{20}$O+p system. $U_1$ and
$F_1$ show strength and shape of the Lane potential in the elastic and
2$^+$-inelastic channels, respectively.} \label{f1}
\end{figure*}

In the present work we have extensively analyzed the elastic and inelastic
$^{18,20}$O+p scattering data at 43 \cite{El00} and $^{20}$O+p data at 30
MeV/nucleon \cite{Je99}. The optical model (OM) analysis was done using the real
folded potential \cite{Kh02} obtained with the density- and isospin dependent
CDM3Y6 interaction \cite{Kh97} and microscopic g.s. densities given by the
Hartree-Fock-Bogoljubov approach \cite{Gr01}. The imaginary optical potential
was parametrized in a Woods-Saxon (WS) form using the CH89 global systematics
\cite{Va91}. Elastic data are well reproduced with the WS strengths slightly
adjusted by OM fit (keeping the same radius and diffuseness given by CH89
systematics) and real folded potential renormalized by a factor $N_R\approx
1.08$ and 1.03 for $^{18}$O and $^{20}$O, respectively. The isospin dependence
of the CDM3Y6 interaction was shown earlier to reproduce the empirical symmetry
energy of asymmetric nuclear matter \cite{Kh96} and it gives also realistic
estimate for the Lane potential $U_1$. In both cases, the ratio of the volume
integrals of $U_1$ and $U_0$ parts of the real (folded) optical potential per
interacting nucleon pair is $J_1/J_0\approx -0.37$, which agrees well with the
observed trend. To illustrate the radial shape of the Lane potential we have
plotted in the left panel of Fig.~\ref{f1} the folded $U_1$ and $U_0$ potentials
for $^{20}$O+p system. An enhancement of $U_1$ strength (approaching around 10
MeV) was found near the surface which must be due to the neutrons in the outer
shell. Since the best-fit $N_R$ factors of the folded potential are quite close
to unity, our result confirms the reliability of the folding model in predicting
the strength and shape of the Lane potential, given a realistic choice for the
effective NN interaction and nuclear g.s. densities.

\begin{table*}[htb]
\caption{Deformation parameters [$\beta_x=\delta_x/(1.2A^{1/3})$] and the
ratios of transition matrix elements for 2$^+_1$ and 3$^-_1$ states in
$^{18,20}$O given by our Folding + DWBA analysis of inelastic $^{18,20}$O+p
scattering data at 30 and 43 MeV/nucleon. $J_1/J_0$ is the ratio of the volume
integrals (per interacting nucleon pair) of $F_1$ and $F_0$ parts of the
inelastic form factor (\ref{e4}).} \label{t1}
\begin{ruledtabular}
\begin{tabular}{c|c|c|c|c|c|c|c|c} 
 \multicolumn{9}{c}{$^{18}$O ($N/Z=1.25,\ \varepsilon=0.11$)} \\ \hline
 $\lambda^\pi$ & $\beta_p$ & $\beta_n$ & $M_n/M_p$ & $\beta_0$ & $\beta_1$ &
 $M_1/M_0$ & $J_1/J_0$ & data \\ \hline
 2$^+$ & $0.331\pm 0.006$ & $0.455\pm 0.023$ & $1.80\pm 0.13$ & $0.401\pm 0.016$ &
 $0.861\pm 0.034$ & $0.286\pm 0.023$ & $-0.992\pm 0.089$ & \cite{El00}\\
 3$^-$ & $0.461\pm 0.003$ & $0.453\pm 0.022$ & $1.35\pm 0.08$ & $0.456\pm 0.014$ &
 $0.432\pm 0.016$ & $0.149\pm 0.010$ & $-0.410\pm 0.029$ & \cite{El00} \\
 \hline
 \multicolumn{9}{c}{$^{20}$O ($N/Z=1.50,\ \varepsilon=0.20$)} \\ \hline
 2$^+$ & $0.250\pm 0.009$ & $0.653\pm 0.032$ & $4.25\pm 0.28$ & $0.500\pm 0.020$ &
 $1.295\pm 0.052$ & $0.619\pm 0.050$ & $-1.258\pm 0.101$ & \cite{El00} \\
 2$^+$ & $0.250\pm 0.009$ & $0.635\pm 0.032$ & $4.13\pm 0.27$ & $0.489\pm 0.019$ &
 $1.248\pm 0.050$ & $0.611\pm 0.050$ & $-1.234\pm 0.099$ & \cite{Je99} \\
 3$^-$ & $0.437\pm 0.003$ & $0.381\pm 0.019$ & $1.55\pm 0.09$ & $0.401\pm 0.012$ &
 $0.308\pm 0.009$ & $0.216\pm 0.013$ & $-0.281\pm 0.017$ & \cite{El00} \\
\end{tabular}
\end{ruledtabular}
\end{table*}

We discuss now the IS and IV strengths of the inelastic $^{18,20}$O+p form
factors. Note that $^{18}$O nucleus is rather well studied and inelastic
$^{18}$O+p data are, therefore, quite helpful in testing the present Folding
approach. By adjusting $M_p$ to the experimental $B(E2\uparrow)=45.1\pm 2.0\
e^2$fm$^4$ \cite{Ra01} and $B(E3\uparrow)=1120\pm 11\ e^2$fm$^6$ \cite{Sp89}
for the first 2$^+$ and 3$^-$ states in $^{18}$O, we obtain $\delta_p=1.040\pm
0.020$ and $1.449\pm 0.008$ fm, respectively. The $E2$ transition strength is
more fragmented in $^{20}$O and the experimental $B(E2\uparrow)=28.1\pm 2.0\
e^2$fm$^4$ \cite{Ra01} for 2$^+_1$ state. There are no $B(E3\uparrow)$ data
available for 3$^-_1$ state in $^{20}$O, and we have assumed a value
$B(E3\uparrow)=1200\pm 12\ e^2$fm$^6$ which was estimated from the experimental
$B(E3\uparrow)$ for 3$^-_1$ state in $^{18}$O using the ratio of the
$B(E3\uparrow)$ values calculated for these two cases in the Quasiparticle
Random Phase Approximation (QRPA) \cite{El00}. As a result, we obtain the
proton deformation lengths $\delta_p=0.815\pm 0.029$ and $1.424\pm 0.008$ fm
for 2$^+_1$ and 3$^-_1$ states in $^{20}$O, respectively. Note that the
numerical uncertainties of the obtained proton deformation lengths are fully
determined by those of the measured $B(E\lambda\uparrow)$ values. Using the
best-fit neutron deformation length from the DWBA analysis of the inelastic
data under consideration, realistic shape of the Lane potential in an inelastic
scattering channel can be obtained. As an example, we have plotted in the right
panel of Fig.~\ref{f1} the 2$^+$-inelastic form factor for $^{20}$O+p system,
where contributions by the IS and IV components are shown explicitly. We
further assign a numerical uncertainty of around 5\% to the deduced neutron
deformation length which gives a cross-section shift within the experimental
errors. The numerical uncertainties of all the deformation parameters and
ratios of transition matrix elements given in Table~\ref{t1} were deduced
directly from those found for the proton and neutron deformation lengths.

Since the CDM3Y6 interaction is \emph{real}, only real nuclear, Coulomb and
spin-orbit transition form factors for $^{18,20}$O are obtained from the folding
calculation \cite{Kh02}. The imaginary nuclear form factor is obtained by
deforming the imaginary part of the optical potential with $\delta_0$ that is
iteratively found from the DWBA fit to the data. Fortunately, nucleon inelastic
scattering at low-to-medium energies is not dominated by the imaginary coupling
\cite{Sa83} and the DWBA cross section is strongly sensitive to the real form
factor which allows an accurate deduction of the (neutron) deformation length.

\begin{figure*}[htb]\vspace*{-2cm}
\begin{minipage}[t]{8.5cm}
\hspace*{-1cm}
\mbox{\epsfig{file=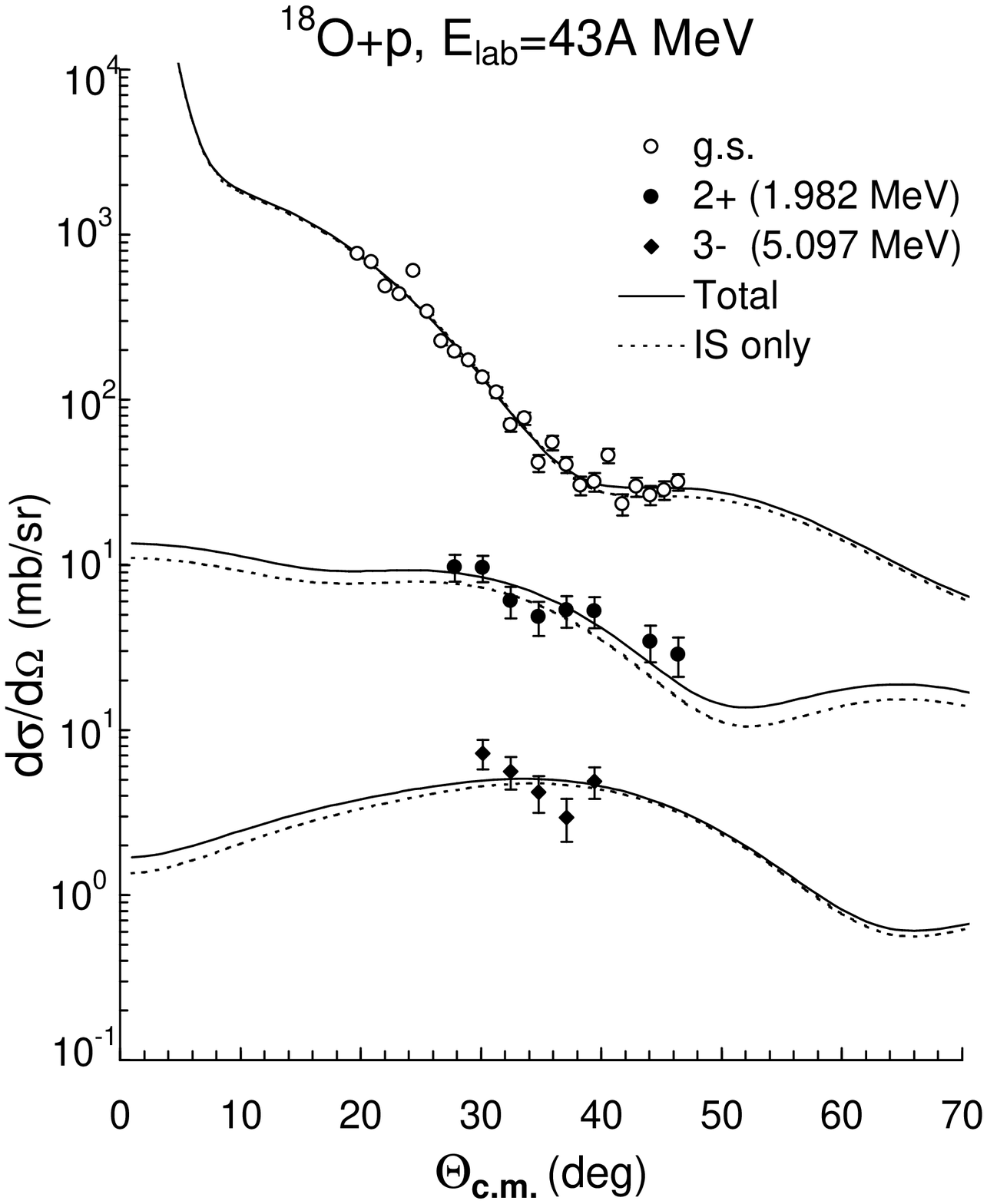,height=10cm}}
\end{minipage}
\hspace{\fill}
\begin{minipage}[t]{8.5cm}
\hspace*{-2cm}
\mbox{\epsfig{file=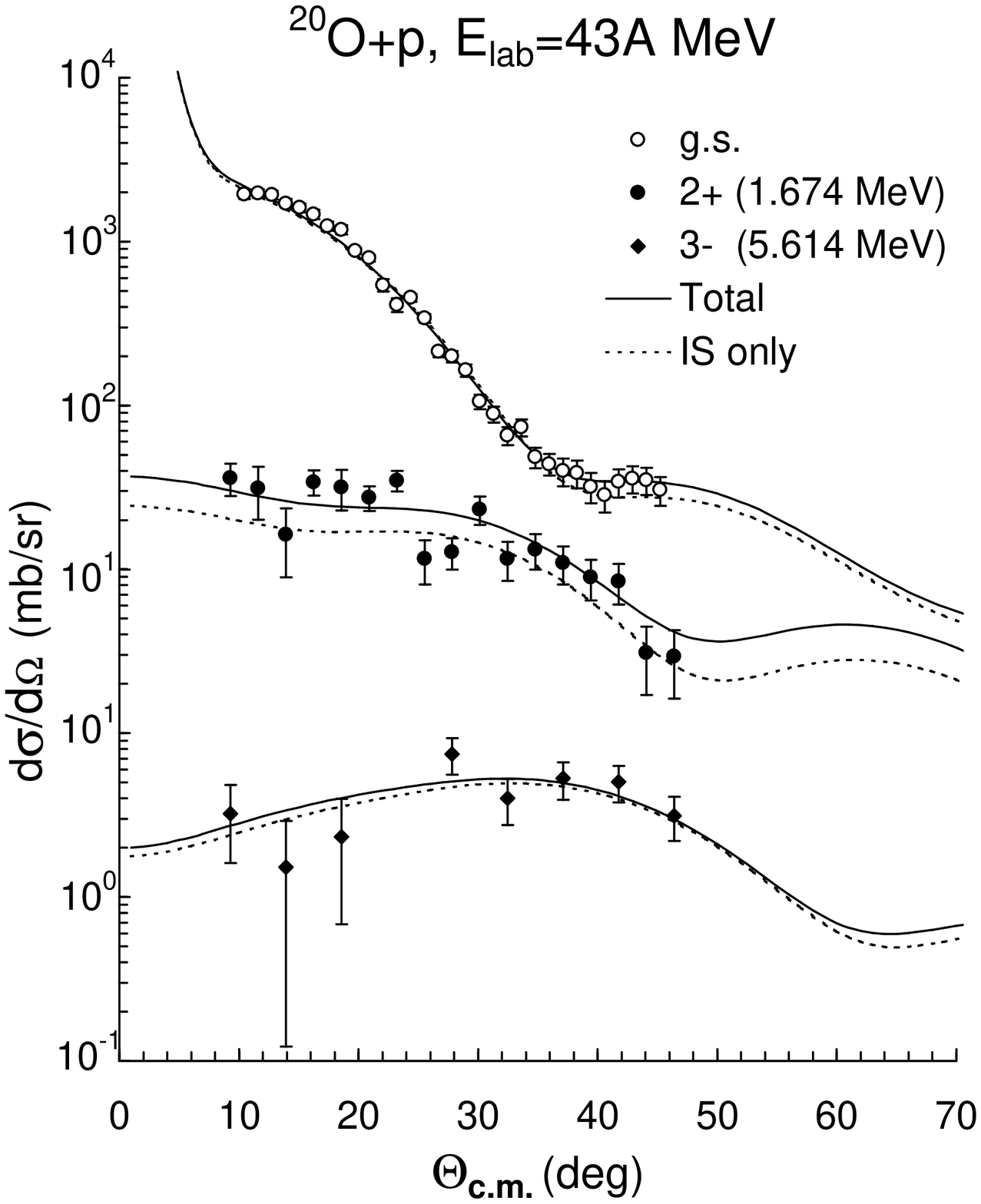,height=10cm}}
\end{minipage}
\vspace*{-2cm} \caption{Elastic and inelastic $^{18,20}$O+p scattering data at
43 MeV/nucleon in comparison with DWBA cross sections given by the folded form
factors. Cross sections given by the isoscalar potentials alone are plotted as
dotted curves.} \label{f2}
\end{figure*}

\begin{figure}[htb]
 \vspace*{-0.3cm}
 \mbox{\epsfig{file=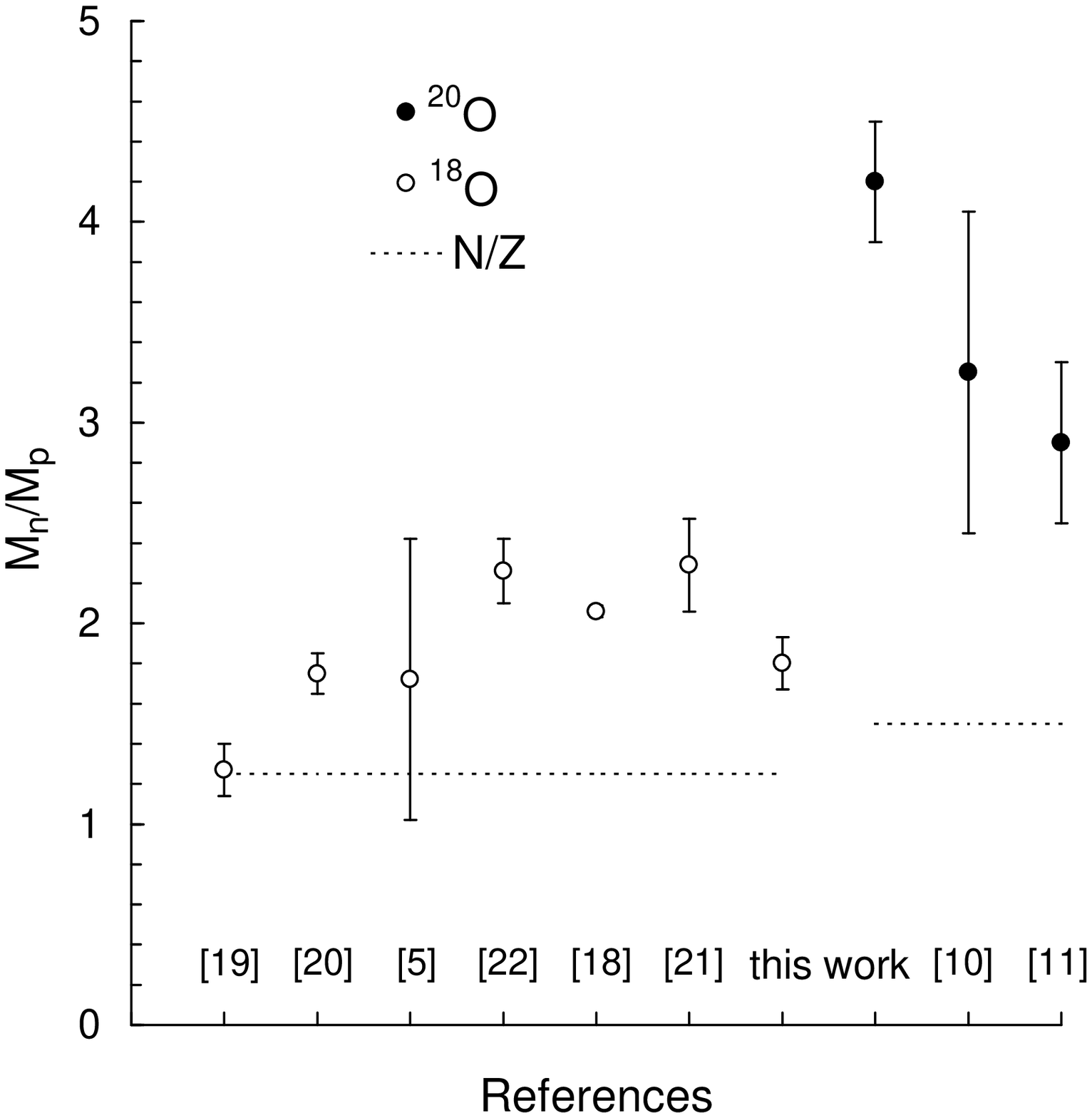,height=10cm}}
\vspace*{-2cm} \caption{$M_n/M_p$ ratios extracted for the lowest 2$^+$
excitations in $^{18,20}$O isotopes.} \label{f3}
\end{figure}

Elastic and inelastic $^{18,20}$O+p cross sections (at 43 MeV/nucleon) obtained
with the best-fit deformation parameters are plotted in Fig.~\ref{f2}. One can
see that the IV contribution is small in the elastic and 3$^-$ inelastic
channels. We found that 3$^-_1$ state is dominantly isoscalar, with the
best-fit $M_n/M_p$ ratio slightly above $N/Z$ and $M_1/M_0$ ratio close to
$\varepsilon$ (see Table~\ref{t1}). This result is well expected because the
3$^-_1$ states in $^{18,20}$O isotopes were shown by the QRPA calculation
\cite{El00} to consist mainly of the (1p$^{-1}_{1/2}$,1d$_{5/2}$) proton
configuration. Strong IV effect was found in 2$^+_1$ inelastic channel.
Structure of 2$^+_1$ state in $^{18}$O has been investigated in numerous
studies like (p,p') reactions at low \cite{Gr80} and intermediate energies
\cite{Ke86} or ($\pi,\pi'$) reactions \cite{Ja78,Iv78,Se88}, and the weighted
average of those results \cite{Je99} gives $M_n/M_p\approx 2$. This value also
agrees fairly with that deduced from a pure isospin-symmetry assumption that
$M_p$ obtained for the mirror $^{18}$Ne nucleus would yield $M_n$ for the
corresponding excited state in $^{18}$O \cite{Be79}. $M_n/M_p$ ratios deduced
from these studies are compared with our result in Fig.~\ref{f3}.

One can see that our result is in a satisfactory agreement with the empirical
data. The obtained IV deformation ($\beta_1\approx 0.86$) is about twice the IS
deformation which indicates a significant IV mixing in this case. Prior to our
work, the only attempt to deduce $\beta_1$ for 2$^+_1$ state in $^{18}$O that we
could find in the literature is the DWBA analysis of (p,p') and (n,n')
scattering data at 24 MeV \cite{Gr80} which gives $\beta_0\approx 0.4$ and
$\beta_1\approx 1.0\pm 0.5$, using prescription (\ref{e2}) and assuming
$\beta_0$ to be the average of the $\beta$ values obtained with (p,p') and
(n,n') data. Although the uncertainty associated with $\beta_1$ is large, this
result agrees reasonably with $\beta_{0(1)}$ given by our analysis.

In contrast to the $^{18}$O case, the (p,p') excitation of 2$^+_1$ state in
radioactive $^{20}$O nucleus has been studied only recently in the inverse
kinematics proton scattering measurements at 43 \cite{El00} and 30 MeV/nucleon
\cite{Je99}. A simple folding analysis using the microscopic QRPA transition
densities and JLM interaction \cite{El00} has failed to fit the data at 43 MeV.
Khan {\sl et al.} obtained $M_n/M_p\approx 1.10\pm 0.24$ and $3.25\pm 0.80$ for
2$^+_1$ excitation in $^{18}$O and $^{20}$O, respectively, after renormalizing
the QRPA densities to the best DWBA fit to the data \cite{El00}. If one uses a
simple (probe-dependent) collective model in the analysis of (p,p') data
\cite{Je99} measured at 24 and 30 MeV for $^{18}$O and $^{20}$O, these values
become $M_n/M_p\approx 1.50\pm 0.17$ and $2.9\pm 0.4$, respectively. Despite the
uncertainty of these data, they do indicate a strong IV mixing in the 2$^+_1$
excitation in $^{20}$O. Our result shows even stronger IV transition strength
for this state and $M_n/M_p$ ratio obtained with the best-fit neutron
deformation length from our folding analysis of 43 MeV \cite{El00} and 30 MeV
data \cite{Je99} is around 4.25 and 4.13, respectively. We adopted, therefore,
an average value of $M_n/M_p\approx 4.2\pm 0.3$ from the values obtained in
these two cases. The IV deformation, given for the first time for 2$^+_1$ state
in $^{20}$O ($\beta_1\approx 1.3$), is nearly three times the IS deformation
($\beta_0\approx 0.5$) and $M_1/M_0\approx 0.6=3\varepsilon$. This leads to a
ratio of the volume integrals of $F_1$ and $F_0$ folded form factors
$J_1/J_0\approx -1.25$ which is significantly higher than that found in the
elastic channel. The relative IV strength in the inelastic 2$^+_1$ channel is,
therefore, $\varepsilon |J_1/J_0|\approx 25\%$ with the IV form factor peaked at
the surface (see Fig.\ref{f1}). This significant contribution by the Lane
potential in the 2$^+_1$ inelastic channel of $^{20}$O+p system amounts up to
40-50\% of the total cross section over the whole angular range as shown in
Fig.\ref{f2}.

In conclusion, a compact folding approach is developed for a consistent study of
strength and shape of the Lane potential in both elastic and inelastic \pA
scattering, and to deduce from the analysis of (p,p') data the IS and IV
deformation parameters which, otherwise, can be deduced only if there are (p,p')
and (n,n') data available at the same energy for the same target. With more data
being measured with the unstable beams, our model should be helpful for the
determination of the isospin distribution in the low-lying excited states of
exotic nuclei which can be used as important `data base' for further nuclear
structure studies. The use of microscopic nuclear densities in our approach
should be encouraged to test the nuclear structure model ingredients by studying
the known excitations and, consequently, to predict the isospin character of
those not yet measured.

The author thanks Paul Cottle, Marcella Grasso, Elias Khan, and Kirby Kemper for
very helpful correspondences. The research was supported, in part, by Natural
Science Council of Vietnam.

\end{document}